\documentclass[prd,amssymb,altaffilletter,runningaddress,superscriptaddress,showpacs,showkeys,epsfig]{revtex4}
\usepackage[dvips]{graphicx}
\usepackage[english]{babel}
\newcommand{\be}{\begin{equation}}
\newcommand{\ee}{\end{equation}}
\newcommand{\ben}{\begin{eqnarray}}
\newcommand{\een}{\end{eqnarray}}

\begin{document}

\title{Chaplygin gas may prevent big trip}

\author{Jos\'e A. Jim\'enez Madrid}
\email{madrid@iaa.es}\affiliation{Colina de los Chopos, Centro de F\'{\i}sica
``Miguel A. Catal\'an'', Instituto de Matem\'aticas y F\'\i sica Fundamental
Consejo Superior de Investigaciones Cient\'\i ficas,
Serrano 121, 28006 Madrid, Spain}
\affiliation{Instituto de Astrof\'\i sica de Andaluc\'\i a, Consejo Superior de Investigaciones Cient\'\i ficas, Camino Bajo de Hu\'etor 50, 18008 Granada, Spain}

\date{\today}

\begin{abstract}
This paper deals with the study of the accretion of a generalized Chaplygin
gas with 
equation of state $p=-A/\rho^\alpha$ onto wormholes. We have 
obtained that when dominant energy condition is violated  the 
size of wormhole increases with the scale factor up to a given plateau.
On the regime where the dominant energy condition is satisfied 
our model predicts a steady decreasing of the wormhole size
as generalized Chaplygin gas is accreted. Our main conclusion is
that the big trip mechanism is prevented in a large region of the
physical parameters of the used model.
\end{abstract}
\pacs{ 04.70.-s, 98.80.-k}
\keywords{accretion, wormholes, dark energy.}
\maketitle

Several astronomical  and cosmological observations,  ranging from
 distant supernovae Ia \cite{Riess:1998cb} to
the cosmic microwave background anisotropy \cite{Spergel:2003cb}, 
indicate  that
the universe is currently undergoing an accelerating stage. It is assumed that
this acceleration is due to some unknown stuff usually dubbed dark
energy, with a positive energy density $\rho>0$ and
with negative pressure $p<-(1/3)\rho.$

There are several candidate models for describing dark energy, being the cosmological constant, $\Lambda$, by far,
the simplest and most popular candidate \cite{Lambda}. Other
interesting models are based on considering a perfect fluid with given
equation of state like in quintessence \cite{quintessence}, K-essence \cite{Armendariz-Picon:1999rj} or 
generalized Chaplygin gas \cite{chaplygin,RBOD,MMG,chaplygin2,MBL-PVM} models.
Note  that
there are also other candidates for dark energy based on brane-world 
scenarios \cite{brane} and modified 4-dimensional Einstein-Hilbert
actions \cite{EHmodified}, where a late time acceleration of the
universe may be achieved, too.

One of the peculiar properties of the resulting cosmological models
is the possibility of occurrence of  a cosmic doomsday, also dubbed big
rip \cite{caldwell}.
The big rip appears in models where dark energy 
particularizes as the so-called phantom energy for which
the dominant energy condition is violated, so that $p+\rho<0$. In these models the scale factor
blows up in a finite time because its cosmic acceleration is even
larger than that  induced by a positive cosmological constant.
In these models every component of the universe  goes beyond the 
horizon of all other universe components in
finite cosmic time. It should be noted that the condition $p+\rho<0$
is not enough for the occurrence of a big rip, i.e., if one 
consider an universe filled with phantom generalized Chaplygin
gas, one can avoid the big rip \cite{Bouhmadi-Lopez:2004me} (see also \cite{McInnes:2001zw}).
Other peculiar properties of phantom energy are that it can make the
exotic substance that fuel wormholes \cite{Lobo:2005us}, triggering 
 the possibility of
occurrence of a big trip \cite{Gonzalez-Diaz:2004vv,Gonzalez-Diaz:2004df,Gonzalez-Diaz:2005yj}, i.e.
if there is a wormhole in an universe filled with phantom energy,
due to processes of phantom energy accretion onto the wormhole, the
size of this wormhole increases in such way that the wormhole
can engulf the universe itself before it reaches the big rip singularity,
at least relative to an asymptotic observer.
Then, the following question arises: Does an universe filled with 
generalized phantom Chaplygin gas which avoids the big rip singularity 
 also escape from the big trip? We have found in the present paper
that for a large range of the Chaplygin parameters no big
trip is predicted by our model, though there still exists 
sufficient room in the physically allowed parameter space to
not exclude the possibility for the occurrence of such a strange
causality disruption.
Thus, generalized Chaplygin gas can be seen to have several 
interesting potential properties,
as it avoids big rip \cite{Bouhmadi-Lopez:2004me}, it
may be used as the stuff to construct wormholes \cite{Lobo:2005vc}, it prevents
the universe to be engulfed by a black hole  \cite{JimenezMadrid:2005rk}, and,
finally it may also circumvent the big trip problem.

We starts by first reviewing the accretion formalism first considered
by Babichev, Dokuchaev and Eroshenko 
\cite{Babichev:2004yx} (see also \cite{Gonzalez-Diaz:2005ex}), generalizing it to the case of wormholes. Throughout this
paper we shall use natural units so that $G=c=1$. 
The  Morris-Thorme static space-time metric of
one wormhole is given by \cite{Morris:1988cz}

\begin{equation}
ds^2=-e^{\Phi(r)}dt^2+\frac{dr^2}{1-\frac{K(r)}{r}}+r^2(d\theta^2+\sin^2\theta d\phi^2),
\end{equation}
where $\Phi(r)$ is the shift function and $K(r)$ is the shape function. We model the 
dark energy in the wormhole
by the test perfect fluid with a negative pressure and  an arbitrary 
equation of state $p(\rho)$, with the energy-momentum tensor

\begin{equation} \label{eq:deftensorenergy}
T_{\mu\nu}=\left(p+\rho\right)u_\mu u_\nu +pg_{\mu\nu},
\end{equation}
where $p$ is the pressure, $\rho$ is the energy density, and $u^\mu=dx^\mu/ds$
is the 4-velocity with $u^\mu u_\mu=-1.$ The zeroth (time) 
component of the energy-momentum conservation law $T^{\mu\nu}{}_{;\nu}=0$ 
can then generally be written as
\begin{eqnarray}   
0=\frac{d}{dr}\left[e^{\Phi(r)}\left(p+\rho\right)\frac{dt}{ds}\frac{dr}{ds}\right]
+e^{\Phi(r)}\left(p+\rho\right)\left[\Phi^\prime(r)+\frac{K^\prime(r)r-K(r)}{2r^2\left(1-\frac{K(r)}{r}\right)}+\frac{2}{r}\right]\frac{dt}{ds}\frac{dr}{ds}. \label{eq:energymomentumconservation}
\end{eqnarray} 

This expression should now be integrated.
The integration of Eq. (\ref{eq:energymomentumconservation}) gives then,
\begin{eqnarray}
&uM^{-2}r^2\left(1-\frac{K(r)}{r}\right)^{-1}\left(p+\rho\right)\sqrt{u^2+1-\frac{K(r)}{r}}=C,\label{eq:energyconservation}
\end{eqnarray}
where $u=dr/ds,$ and $M$ is the exotic mass of the wormhole which,
following the procedure of Ref.  \cite{Babichev:2004yx},
has been introduced to render the integration constant $C$ to have
the dimensions of an energy density (note that we are using
natural units), and, without any loss of generality for
our present purposes, we have adhered to the case where $\Phi^\prime=0$.

Another integral of motion can be derived by using the projection of the 
conservation law for energy-momentum tensor along the four-velocity, i.e. the flow equation $u_\mu T^{\mu\nu}{}_{;\nu}=0$. For a perfect fluid, this 
equation reduces to
\begin{equation}
u^\mu\rho_{,\mu}+\left(p+\rho\right)u^\mu{}_{;\mu}=0. \label{eq:energyvelocityconservation}
\end{equation}
The integration of Eq. (\ref{eq:energyvelocityconservation}) gives the
second integral of motion that we shall use in what follows

\begin{equation}\label{eq:energyfluxequation}
M^{-2}r^2 u\left(1-\frac{K(r)}{r}\right)^{-1/2}e^{\int^\rho_{\rho_\infty}\frac{d\rho}{p+\rho}}=-A,
\end{equation}
where $u<0$ in the case of a fluid flow directed toward the wormhole, and
$A$ is a positive dimensionless constant. 
Eq.~(\ref{eq:energyfluxequation}) gives us the energy flow induced
in the accretion process. From Eqs.~(\ref{eq:energyconservation}) and (\ref{eq:energyfluxequation}) one can easily get

\begin{eqnarray}
\left(p+\rho\right)\left(1-\frac{K(r)}{r}\right)^{-1/2}\sqrt{u^2+1-\frac{K(r)}{r}} \ e^{-\int^\rho_{\rho_\infty}\frac{d\rho}{p+\rho}}=C_{2}, \label{eq:final}
\end{eqnarray}
where $C_{2}=-C/A=\tilde{A}\left(p(\rho_\infty)+\rho_\infty\right)$, with $\tilde{A}$ a positive constant.

The rate of change of the exotic mass of wormhole due to accretion 
of dark energy can be derived by integrating over the surface area 
the density of momentum $T_0 {}^r$, that is \cite{Landau}  

\begin{equation} \label{eq:rategeneral}
\dot{M}=-\int T_0 {}^r dS, 
\end{equation}
with $dS=r^2\sin\theta d\theta d\phi,$. Using Eqs. (\ref{eq:deftensorenergy}), (\ref{eq:energyfluxequation}) and 
(\ref{eq:final}) this can be rewritten as \cite{Gonzalez-Diaz:2005yj}
\begin{equation} \label{eq:rate}
\dot{M}=-4\pi D M^2\sqrt{1-\frac{K(r)}{r}}\left[p\left(\rho_\infty\right)+\rho_\infty\right],
\end{equation}
with the constant $D=A\tilde{A}>0$. For the relevant asymptotic regime $r\rightarrow\infty$ where the big trip occurs, the rate $\dot{M}$ reduces to

\begin{equation}\label{eq:ratem}
\dot{M}=-4\pi M^2D\left(p+\rho\right).
\end{equation}
We see then that the rate for the wormhole exotic mass due to accretion of dark
energy becomes exactly the negative to the similar rate in the case of
a Schwarzschild black hole, asymptotically.
For current quintessence models, the use of a scale factor $R=R_0\left(1+\sqrt{6\pi\rho_0}(1+w)(t-t_0)\right)^{2/[3(1+w)]}$ for
$w<-1$ (which corresponds to the solution of the general equation $-3H(1+w)=2\dot{H}/H,$ with $\dot{ }=d/dt$ and $H=\dot{R}/R$ for $w$=constant \cite{EscalaFRW}), gives arise to a singular behaviour of the size of
the wormhole before reaching the big rip, that is the big trip.

We shall derive now the expression for the rate $\dot{M}$ in the case of a generalized Chaplygin gas. This can be described as a perfect fluid with the equation of state \cite{chaplygin} 
\begin{equation}\label{eq:stateChaplygin}
p=-A_{ch}/\rho^\alpha,
\end{equation}
where $A_{ch}$ is a positive constant and $\alpha>-1$ is a parameter. In the particular case $\alpha=1,$ the equation of state (\ref{eq:stateChaplygin}) corresponds to a Chaplygin gas. The conservation of the energy-momentum tensor implies

\begin{equation}
\rho=\left(A_{ch}+\frac{B}{R^{3\left(1+\alpha\right)}}\right)^{1/\left(1+\alpha\right)},
\end{equation}
with $B\equiv \left( \rho_0^{\alpha +1} -A_{ch}\right)R_0^{3\left(\alpha+1\right)}$, $R\equiv R(t)$ is the scale factor and the subscript ``0'' means initial value. Now, from the Friedmann equation we can get

\begin{equation}\label{eq:rater}
\dot{R}=\sqrt{\frac{8\pi}{3}}R\left(A_{ch}+\frac{B}{R^{3\left(1+\alpha\right)}}\right)^{1/\left[2\left(1+\alpha\right)\right]}.
\end{equation}

Hence, from Eqs. (\ref{eq:ratem})-(\ref{eq:rater}) we obtain

\begin{equation}
M=\frac{M_0}{1-DM_0\sqrt{\frac{8\pi}{3}}\left[\rho^\frac{1}{2}-\rho_0^{1/2}\right]},
\end{equation}
For the case where the dominant energy condition is
preserved, i.e. $B>0$, we obtain that $M$
decreases with time and  tends to a constant value. On the other hand,
$M$ is seen to decrease more rapidly as parameter $\alpha$ is
made smaller. 
If the dominant energy condition is assumed to be violated, i.e. $B<0$, as
phantom energy is assumed to require \cite{Bouhmadi-Lopez:2004me},
then $M$  increases with time, with $M$ tending to maximum, nonzero constant values. Making $|B|$ or $\alpha$ smaller, makes the evolution quicker.

When time goes to infinity, then the exotic mass of wormhole approaches
to

\begin{equation}\label{eq:massfinal}
M=\frac{M_0}{1-DM_0\sqrt{\frac{8\pi}{3}}\left(A_{ch}^\frac{1}{2(1+\alpha)}-\rho_0^{1/2}\right) },
\end{equation}
that is a generally finite value both for $B>0$ and $B<0$. 
Thus, at first sight it could be thought that, unlike
what happens in phantom quintessence models, the presence of a generalized
Chaplygin gas precludes the eventual occurrence of the big trip
phenomenon. However, such a conclusion cannot be guaranteed as
the size of the wormhole throat could still exceed the
size of the universe during its previous evolution. Note that for e.g. wormholes with zero tidal force, one can consider that the exotic matter is confined into an arbitrarily small region around the wormhole throat, then the radius of wormhole throat becomes roughly proportional to its mass. That is to say,
the question remains whether the wormhole would grow eventually 
rapidly enough or not to engulf the universe during the evolution to its final classically stationary state.
That question should be settled down before reaching a conclusion 
on the possibility of the big trip in these models. Actually, in order for avoiding a big trip, the following two conditions
are also required: (i) $R\not=M$ along the entire evolution, and (ii) that
$N(R)=\dot{R}/\dot{M}$ be always an increasing function along that evolution. The
first of these conditions implies that the
function
\begin{equation}
f(M)=M-MM_0 D \sqrt{\frac{8\pi}{3}}\left[\left(A_{ch}+\frac{B}{M^{3(1+\alpha)}}\right)^\frac{1}{2(1+\alpha)}-\rho_0^{1/2}\right]-M_0,
\end{equation} 
be nonvanishing everywhere. We analyze this question by taking the zeros 
of the second derivative
\begin{equation}\label{eq:zeros}
f^{\prime\prime}(M)\equiv\frac{d^2 f(M)}{dM^2}=\sqrt{6\pi}DB\frac{M_0}{M^{3(1+\alpha)+1}}\left(A_{ch}+\frac{B}{M^{3(1+\alpha)}}\right)^\frac{-2\alpha-1}{2(1+\alpha)}\left[1-3\left(1+\alpha\right)+\frac{3B(2\alpha+1)}{2M^{3(1+\alpha)}}\left(A_{ch}+\frac{B}{M^{3(1+\alpha)}}\right)^{-1}\right].
\end{equation}
Now, by taking into account that physically,
\begin{equation}\label{eq:condition}
M>M_0>\left(-\frac{B}{A_{ch}}\right)^\frac{1}{3(1+\alpha)},
\end{equation}
The second inequality meaning that, during its evolution, the radio of
the universe should be larger than its initial size.
Eq.~(\ref{eq:zeros}) can be reduced to imply for the zeros 
\begin{equation}
1-3\left(1+\alpha\right)+\frac{3B(2\alpha+1)}{2M^{3(1+\alpha)}}\left(A_{ch}+\frac{B}{M^{3(1+\alpha)}}\right)^{-1}=0,
\end{equation} 
whose solution would read $M^{3(1+\alpha)}=-B/[2(2+3\alpha)A_{ch}]$.
Nevertheless, from condition~(\ref{eq:condition}) we have that $0<2(2+3\alpha)<1$ which, in turns, implies that there could be at most three crossing
points, as this interval allows for values of $\alpha$ within the range of the
generalized Chaplygin gas models, i.e. $\alpha>-1$. For $\alpha$-values outside the
interval $0<2(2+3\alpha)<1$, but still inside the Chaplygin range, one might expect just two points at most. It follows from this analysis that
a big trip could in principle happens.

As to condition (ii), we obtain from Eqs.~(\ref{eq:ratem}) and (\ref{eq:rater})
that 
\begin{equation}\label{eq:g}
N(R)=-\frac{1}{BM_0^2 D\sqrt{6\pi}}R^{3(1+\alpha)+1}\rho^\frac{2\alpha+1}{2}\left\{1-M_oD\sqrt{\frac{8\pi}{3}}\left[\rho^\frac{1}{2}-\rho_0^{1/2}\right]\right\}^2.
\end{equation}
From Eq.~(\ref{eq:g}) it follows that $N(R_0)\geq 1$ because the size of
the wormhole must be quite smaller than that of the universe initially. Thus, an
always increasing $N(R)$ should necessarily imply that the scale factor increased more
rapidly than $M$ did, so preventing any big trip to occur. 
By differentiating $N(R)$ with respect to $R$ and taking into account
that $B<0$ for Chaplygin phantom, it can be checked that $N(R)/dR>0$ in the
general case that $\alpha$ does not reach values sufficiently close to $-1$;
that is, it follows from Eq.~(\ref{eq:massfinal}) that inside the interval
\begin{equation}\label{eq:alphabigtrip}
-1<\alpha<\frac{\ln A_{ch}}{\ln\left(\sqrt{\frac{3}{8\pi}}\frac{1}{M_0 D}+\rho_0^{1/2}\right)^2}-1,
\end{equation} 
a big trip would still take place.

On the other hand, the question still remains on what happens with the 
grown-up wormhole once it has
reached its maximum, final size. Since the wormhole size tends to become
constant at the final stages of its evolution and it is
rather a macroscopic object, it would be subjected to
chronology protection \cite{Hawking:1991nk}. In fact,
one expected that vacuum polarization created particles which 
catastrophically accumulated on the chronology
horizon of the wormhole making the corresponding renormalized
stress-energy tensor to diverge and hence the wormhole 
would disappear.

It would be quite interesting to probe the region of parameter
space $(\alpha,A_{ch},H_0,\Omega_K,\Omega_\phi)$ allowed by current
observations in order to determine whether there exist any allowed
sections leading to a big trip. However, all available analysis 
\cite{Bertolami:2004ic,Biesiada:2004td,Zhu:2004aq,Colistete:2005yx}
are restricted to the physical region where no dominant energy condition 
is violated. Therefore, the section described by the interval
implied by Eq.~(\ref{eq:alphabigtrip}) necessarily is outside 
the analyzed regions. One had to extend the investigated domains to include values of
parameter $A_{ch}>1$ to probe the parameters space relative to the big trip. In
any events the range of $\alpha$ values compatible with a big trip
has been seen to be extremely narrow and hence the occurrence of
the big trip phenomenon in the generalized Chaplygin gas model
appears to be highly unlikely.  
 
In this paper we have studied the accretion of a generalized
Chaplygin gas
onto a wormhole.
First, we have reviewed the accretion formalism originally considered by  
 Babichev, Dokuchaev and Eroshenko 
\cite{Babichev:2004yx} for the case of a wormhole \cite{Gonzalez-Diaz:2005yj}.
We have then
applied 
such a formalism to the generalized Chaplygin gas model.  
The evolution of exotic mass with the 
accretion of Chaplygin dark energy has been 
first considered for the case that the dominant energy condition
is satisfied. It has been seen
that in that case the mass decreases with cosmic time.

If accretion involves Chaplygin phantom energy, then $M$ increases
from its initial value, tending to reach a plateau as cosmic time
goes to infinity. 
It is obtained that for a wide region of the Chaplygin parameters no
big trip is predicted, contrary to what happens in quintessence and
K-essence dark energy models. However, as far as the Chaplygin
regime tends to match the quintessence regime, but presumably still
within the Chaplygin region, the possibility for a big trip at a finite
time in the future is not excluded. Finally, we also argued that the fate
of the final wormhole is to be destabilized by
quantum vacuum processes. To conclude, the generalized Chaplygin gas has several  very interesting features 
and may circumvent several singularities that appear
in the usual quintessence models.
Whether or not the above features can be taken to imply that
Chaplygin gas is a more consistent component than usual quintessential or K-essence dark
energy component with $w<-1$ is a matter that will depend on both the
intrinsic consistency of the models and the current observational 
data and those that can be expected in the future.

\section*{Acknowledgements}

\noindent We acknowledge P.F. Gonz\'alez-D\'{\i}az, Mar\'{\i}a del Prado Mart\'{\i}n 
and Milagros Rodr\'{\i}guez for constructive discussions and criticisms. This work was
supported by DGICYT under Research Projects BMF2002-03758 and
BFM2002-00778. 

\end{document}